\documentclass[journal]{IEEEtran}

\usepackage{ifpdf}
 \ifpdf
     \usepackage[pdftex]{graphicx}
 \else
     \usepackage[dvips]{graphicx}
 \fi

\graphicspath{{figures/}}
 \usepackage{multirow}

\usepackage{cite}
\usepackage[caption=false]{subfig}

\usepackage{epstopdf}					
\usepackage[keeplastbox]{flushend}

\usepackage[cmex10]{amsmath}
\usepackage{amssymb}

\usepackage{xcolor}
\usepackage{mdwmath}
\usepackage{mdwtab}

\begin{document}

\title{Modulation Classification Using Received Signal's Amplitude Distribution for Coherent Receivers}
\author{Xiang Lin, Yahia A. Eldemerdash, Octavia A. Dobre, Shu Zhang and Cheng Li
\thanks{The authors are with the Faculty of Engineering and Applied Science, Memorial
University, St. John's, NL, A1B 3X5, Canada (e-mail: xiang.lin@mun.ca).}}

\maketitle

\begin{abstract}
In this letter, we propose a modulation classification algorithm which is based on the received signal's amplitude for coherent optical receivers. The proposed algorithm classifies the modulation format from several possible candidates by differentiating the cumulative distribution function (CDF) curves of their normalized amplitudes. 
The candidate with the most similar CDF to the received signal is selected. The measure of similarity is the minimum average distance between these CDFs. Five commonly used quadrature amplitude modulation formats in digital coherent optical systems are employed. Optical back-to-back experiments and extended simulations are carried out to investigate the performance of the proposed algorithm. Results show that the proposed algorithm achieves accurate classification at optical signal-to-noise ratios of interest. Furthermore, it does not require carrier recovery. 
\end{abstract}

\begin{IEEEkeywords}
Elastic optical communication systems, Modulation classification, Cumulative distribution function.
\end{IEEEkeywords}
\IEEEpeerreviewmaketitle


\section{Introduction}
The modern optical fiber communication networks are faced with new challenges due to the significant request for bandwidth. 
Furthermore, the fluctuating nature of the network traffic reflects the inefficiency of the fixed spectrum grid in current dense wavelength division multiplexing  systems. Recently, the elastic optical networks have been considered to deal with this dilemma by maximizing spectral and energy efficiencies  
 \cite{gerstel2012elastic}. Multiple parameters such as the modulation format, bit rate, and channel spacing can be changed dynamically to achieve flexibility for different scenarios. As most digital coherent receivers are adopting modulation format-dependent signal processing algorithms, as well as due to the dynamic change of the modulation format, automatic modulation classification (MC) becomes an essential part of re-configurable coherent receivers. 

Research on MC in wireless communications has been carried out for decades \cite{Dobre2015signal,dobre2007survey,dobre2005blind}. In coherent optical communications, a number of techniques have been proposed in the most recent years. For example, machine learning algorithms are applied to recognize signal's amplitude histograms \cite{khan2016modulation} or the Stokes space-based signal representation \cite{borkowski2013stokes}; however, they require either prior training or iterative processing. Image processing techniques like the connected component analysis are employed in \cite{bo2016modulation} to classify the modulation format in the Stokes space domain, but such features are appropriate only for lower-order modulation formats. A non-iterative clustering algorithm is proposed in \cite{mai2017stokes}; however, tracking the state of polarization and recovering the initial polarization state are required before the MC stage. 
Algorithms relying on the received signal's power distribution have been also employed for optical MC. The peak-to-average power ratio is chosen as a decision metric to distinguish different modulation formats in \cite{bilal2015blind}; however, different thresholds need to be established for each optical signal-to-noise ratio (OSNR). Certain ratios  obtained from the normalized power distribution are selected to characterize the modulation formats in \cite{liu2014modulation}; nevertheless, selection of these ratios is necessary for different candidate modulation formats. 

\begin{figure}[t]
\centering{}\includegraphics[width=0.8\columnwidth]{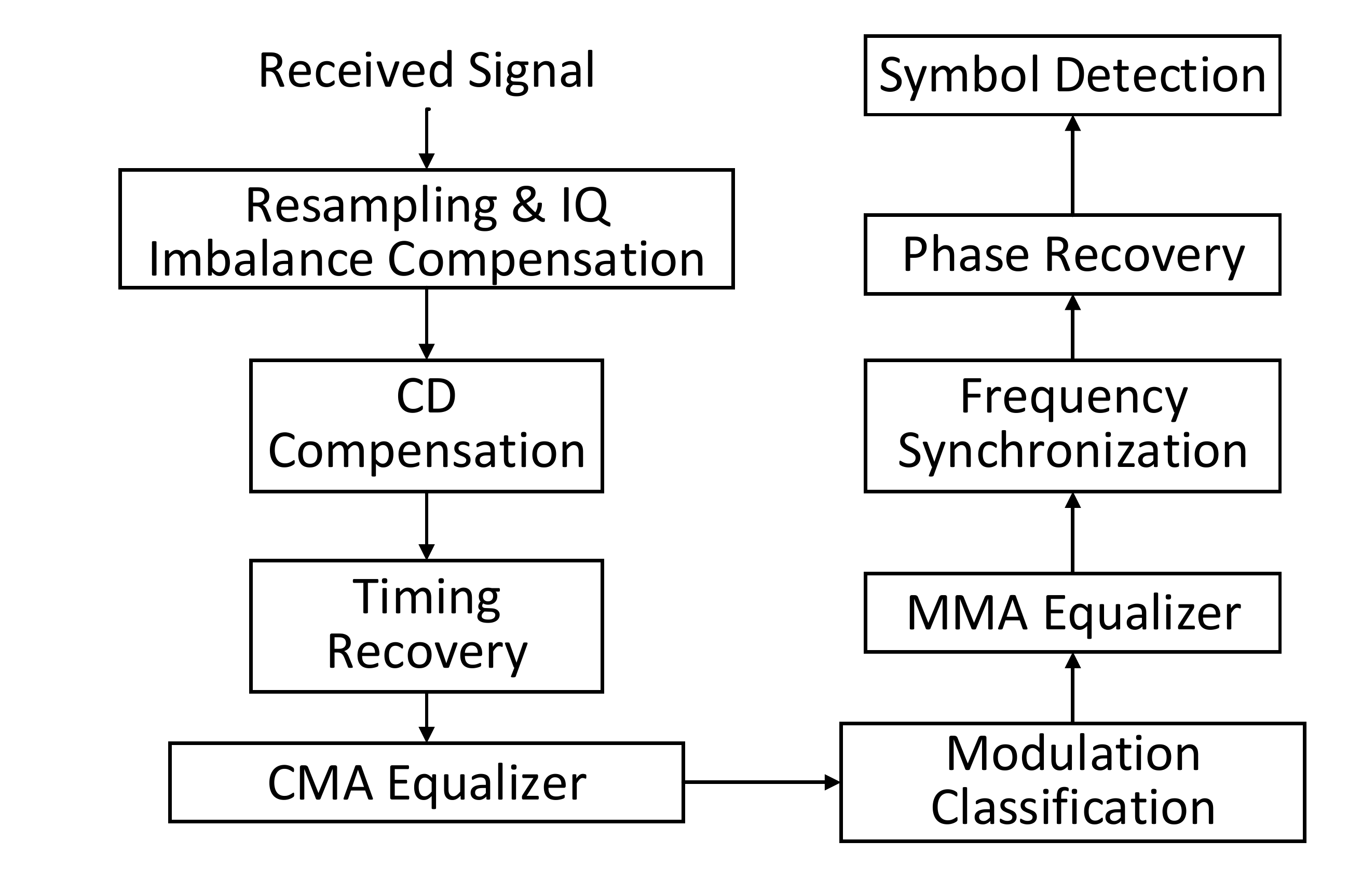}\caption{Structure of the digital receiver.}
\setlength{\belowcaptionskip}{-6.cm}
\end{figure}
\setlength\textfloatsep{-0.02cm}

In this letter, we propose an MC algorithm that requires small number of samples and can be implemented before carrier recovery. The discriminating feature is the cumulative distribution function (CDF) of the received signal's normalized amplitude \cite{wang2016fold} which is obtained after the constant modulus algorithm (CMA) equalization in the digital receiver. The CDF of the normalized amplitude of the received signal is compared with the reference CDFs of all possible candidate modulation formats. The candidate with the closest average distance to the received signal's CDF is chosen. Since $M$-ary QAM is most commonly used in optical communications because of its high spectral efficiency \cite{winzer2012high}, 4-QAM, 8-QAM, 16-QAM, 32-QAM and 64-QAM polarization-division multiplexing (PDM) modulation formats are considered in our experiments and simulations to evaluate the performance of the proposed algorithm. Results show that the proposed algorithm achieves a very good performance with a reduced number of samples. 



\section{Proposed MC Algorithm}

\subsection{Digital Receiver Structure and Discriminating Feature}

The schematic diagram of a digital receiver including the MC module is shown in Fig. 1. The in-phase-quadrature (IQ) imbalance and chromatic dispersion (CD) are compensated first. Then, the CMA is applied to equalize the polarization mode dispersion (PMD). The proposed MC algorithm follows, employing the CDF of signal's normalized amplitudes. After the MC module, the multi-modulus algorithm (MMA), frequency synchronization and phase recovery are applied sequentially. The MC algorithm is signal's amplitude-based, and thus, it does not require carrier recovery. On the other hand, knowledge of the modulation format is beneficial for the following digital processing steps, such as frequency offset and phase noise compensation. 
The pattern differences among the five modulation format constellations can be converted into the one-dimensional CDF of the signal's normalized amplitudes. The CDFs for different QAM formats are shown in Fig. 2, where the amplitude is normalized to its mean. In order to highlight the discriminating feature, OSNR = 30 dB is considered.  These recognizable curves represent the basis of the proposed MC algorithm.

\begin{figure}
\begin{centering}
\includegraphics[width=0.75\columnwidth]{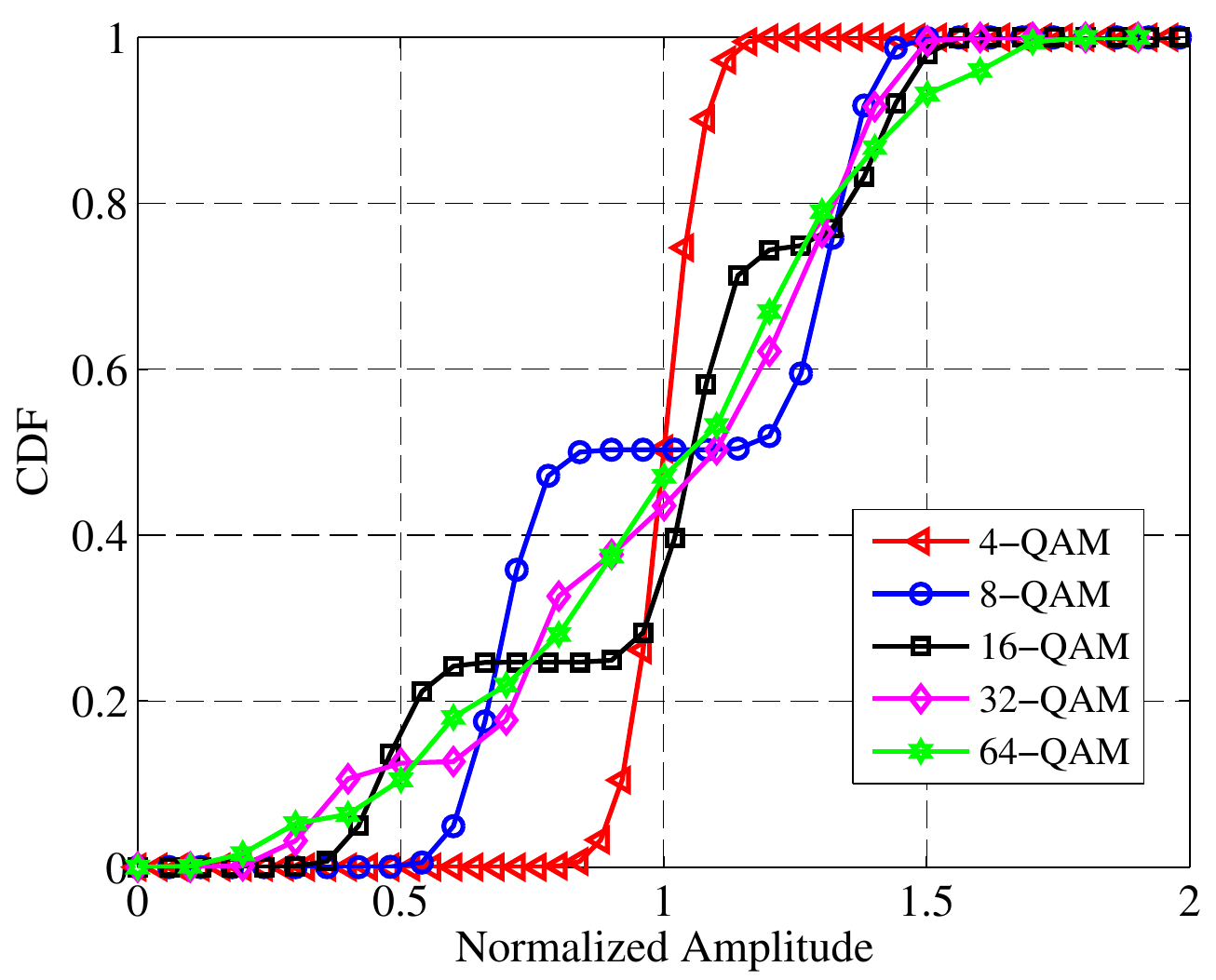}
\par\end{centering}
\caption{CDFs for $M$-ary QAM signals for OSNR = 30 dB.}
\end{figure}

\subsection{CDF-based MC Algorithm}

Before MC is performed, we assume that CD is compensated and the two polarizations are separated. Therefore, the additive white Gaussian noise (AWGN) corrupts the signal primarily. Similar to \cite{bilal2015blind}, we assume that the OSNR of the received signal is known. Then, the receiver is able to emulate the signals' CDFs of all possible candidate modulation formats with the known OSNR. The reference signal's samples can be expressed as:
\begin{equation}
y_{m,k}=x_{m,k}+n_{k},  1 \le k \le K,
\end{equation}
where $y_{m,k}$ and $x_{m,k}$ is the $k$th symbol of the received and transmitted signal with the $m$th modulation format, respectively. Here we have five candidates:  $m=1$ is 4-QAM; $m=2$ is 8-QAM; $m=3$ is 16-QAM; $m=4$ is 32-QAM and $m=5$ is 64-QAM. $n_{k}$ is AWGN with variance $\sigma^{2}$. $K$ is the number of samples. 
To add the corresponding noise, the OSNR should be converted into SNR according to \cite{ives2011estimating}:
\vspace{-0.1cm}
\begin{equation}
{\rm SNR(dB)=OSNR(dB)-10log_{10}}(R_{s}/B_{ref}),
\end{equation}
where $B_{ref}=0.1$ nm ($\approx $ 12.5 GHz at 1550 nm) and $R_{s}$ is the symbol rate. Once all possible signals are generated, we can obtain their CDFs as references. Then, comparisons are carried out between the received signal's CDF and each reference CDF. The one with the most similar CDF shape to the received signal's CDF is decided as the transmitted modulation format.

To measure this similarity, the average distances between the received signal's CDF and each candidate reference CDF are calculated. In (3), $F_{1,m}(z_{k})$ is the CDF of the $m$th candidate, and the received signal's CDF is represented by $F_{0}(z_{k})$. Then, the average distance is calculated by (4), where $\mu_{m}$ is the average distance between the received signal's CDF and the $m$th candidate CDF. The minimum average distance indicates the most likely transmitted modulation format, according to (5).
\begin{equation}
F_{m}(z_{k})=\left|F_{1,m}(z_{k})-F_{0}(z_{k})\right|,1\le k \le K,
\end{equation}
\begin{equation}
\mu_{m}=K^{-1}\sum\limits_{k=1}^{K} F_{m}(z_{k}),
\end{equation}
\begin{equation}
\hat{m}=\arg \min\limits_{m}(\mu_{m}).
\end{equation}

The amplitude-based feature of the proposed algorithm is appealing because it tolerates the constellation rotation caused by the frequency offset and carrier phase noise. This enables the application of MC at an earlier stage in the digital coherent receiver. However, the similarity of the CDFs for different modulation formats increases when the OSNR decreases. Therefore, investigations are needed to verify the MC performance in the OSNR range of practical interest.

\vspace{-0.25cm}
\section{Experimental Results}

We implement the proposed algorithm in optical back-to-back experiments. The setup is shown in Fig. 3. The arbitrary waveform generator (AWG) operates at 12.5 GBd. The sampling rate of the oscilloscope is 50 Gsamples/s. Two free running lasers with 1550.12 nm wavelength are acting as transmit laser and local oscillator, respectively. The linewidth of each laser is about 10 kHz, and the frequency offset is around 200 MHz. An erdium-doped fiber amplifier (EDFA) with fixed output power and a variable attenuator (VOA) are combined to adjust the OSNR values from 10 dB to 20 dB. An optical bandpass filter (OBPF) with 0.6 nm bandwidth is placed after the EDFA. The power entering the integrated coherent receiver (ICR) is around -2 dBm, while the output power of the local oscillator is 13.5 dBm. We measure the OSNR with the optical spectrum analyser (OSA). The data is collected from the oscilloscope and processed offline. After the resampling and the IQ imbalance compensation modules, a symbol-spaced butterfly-type adaptive filter is employed to separate the two polarizations. During the MC stage, similar to \cite{liu2014modulation}, 10000 samples of the received signal are collected to derive the empirical CDF of the normalized amplitude. 

\begin{figure}   
\begin{centering}
\includegraphics[width=0.95\columnwidth]{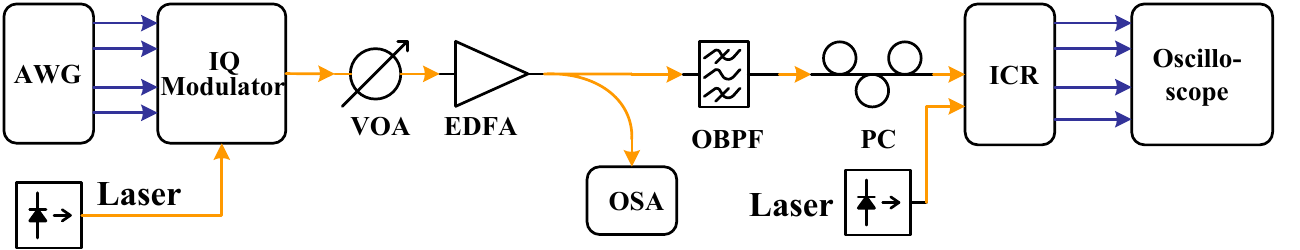}
\par\end{centering}
\caption{Experimental back-to-back system setup. PC: polarization controller.}
\setlength{\belowcaptionskip}{-2.cm}
\end{figure}

\begin{figure}[t]   
\begin{centering}
\includegraphics[width=0.8\columnwidth]{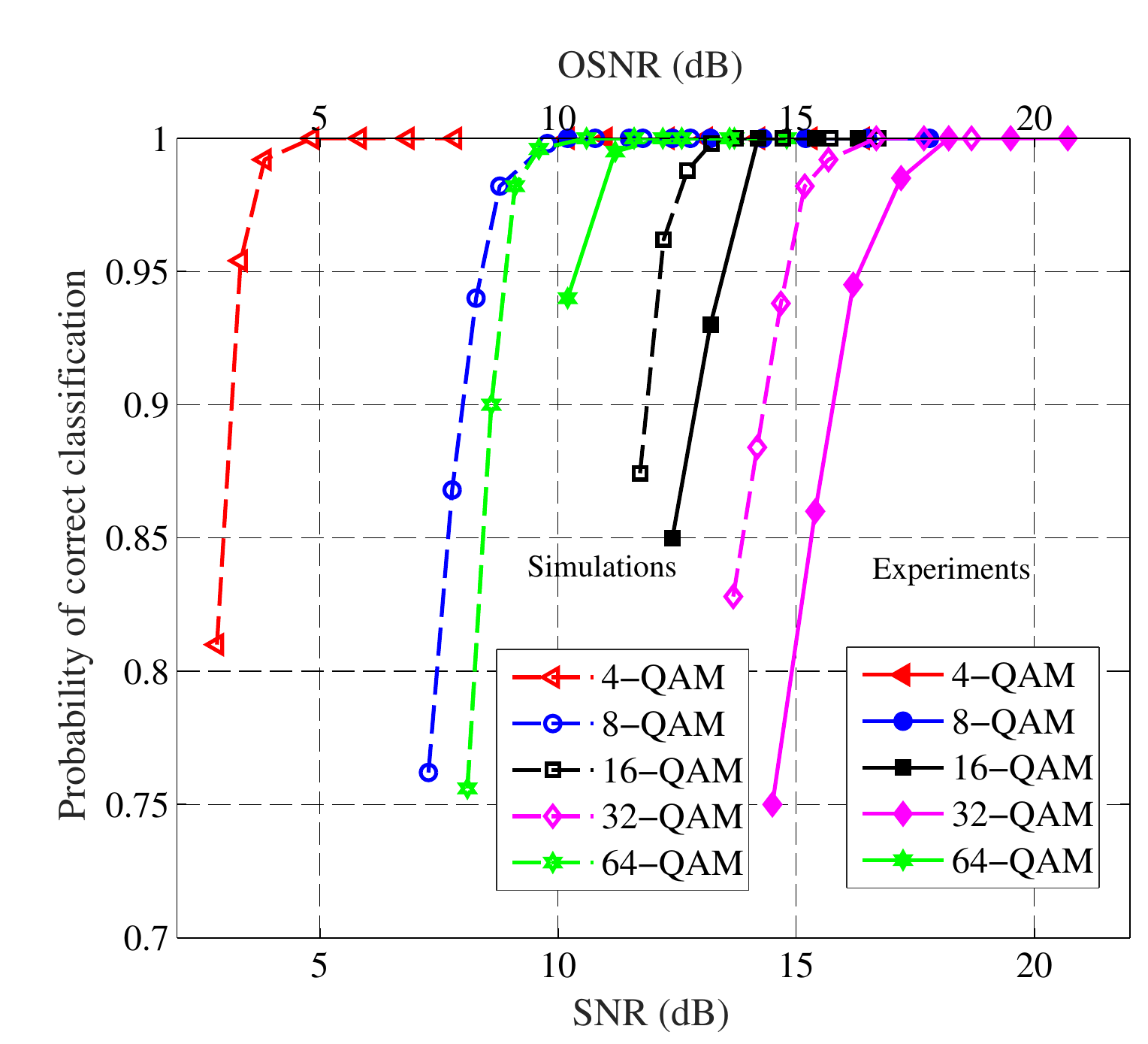}
\par\end{centering}
\caption{Probability of correct classification vs. SNR (bottom) and OSNR (top). }
\end{figure}

The probability of correct classification obtained from 200 realizations for each modulation format is shown in Fig. 4. 
The simulation results of back-to-back are also shown in this figure. We can see that the experimental results have around 1.5 dB penalty compared to simulation results for 16-QAM, 32-QAM and 64-QAM cases. This is mainly caused by the additional electrical noise from imperfect devices such as the ICR, as well as the OSNR measurement error from the OSA.
Note that 64-QAM can be detected with lower OSNR than 16-QAM and 32-QAM; this can be explained as follows. From Fig. 2, it can be seen that the CDFs of 16-QAM, 32-QAM and 64-QAM are close to each other. When the noise level increases, the received signal's CDF curve corresponding to 16-QAM and 32-QAM is likely to be decided as 64-QAM.

\section{Further Investigations}

In order to extend our investigation with optical fiber and higher data rate, we use the VPItransmissionMaker 9.7 to carry out simulations. The system architecture is shown in Fig. 5. The transmitter generates  PDM 4-QAM, 8-QAM, 16-QAM, 32-QAM and 64-QAM signal at 32 GBd rate.  The laser operates at 1550.12 nm with a 100 kHz linewidth. Assuming that perfect IQ modulators are applied, the optical signal is then fed into a 800 km long optical fiber link. After that, the received signal is filtered by an optical low-pass filter with 0.6 nm bandwidth followed by the coherent detector. A local laser is set with 100 kHz linewidth and 200 MHz frequency offset from the transmitted laser.

To simplify the simulations, we ignore the fiber non-linearity first. 500 independent simulations with different noise seeds are carried out for each OSNR.
Fig. 6 illustrates the required SNR and OSNR when the successful classification rate is larger than 0.75. The OSNR ranges of different modulation formats are chosen based on pratical operation of coherent optical systems. The SD-FEC thresholds for 28 GBd rate PDM system with 4-QAM, 8-QAM, 16-QAM, 32-QAM and 64-QAM are illustratd by the vertical dash lines \cite{mai2017stokes}. It can be seen from Fig. 6 that the proposed algorithm can achieve 100\% successful classification rate with lower OSNR values than the SD-FEC thresholds. According to Figs. 4 and 6, the simulation results show the same tendency of  correct classification rate as the experimental results.

\begin{figure}
\centering{}\includegraphics[width=0.8\columnwidth]{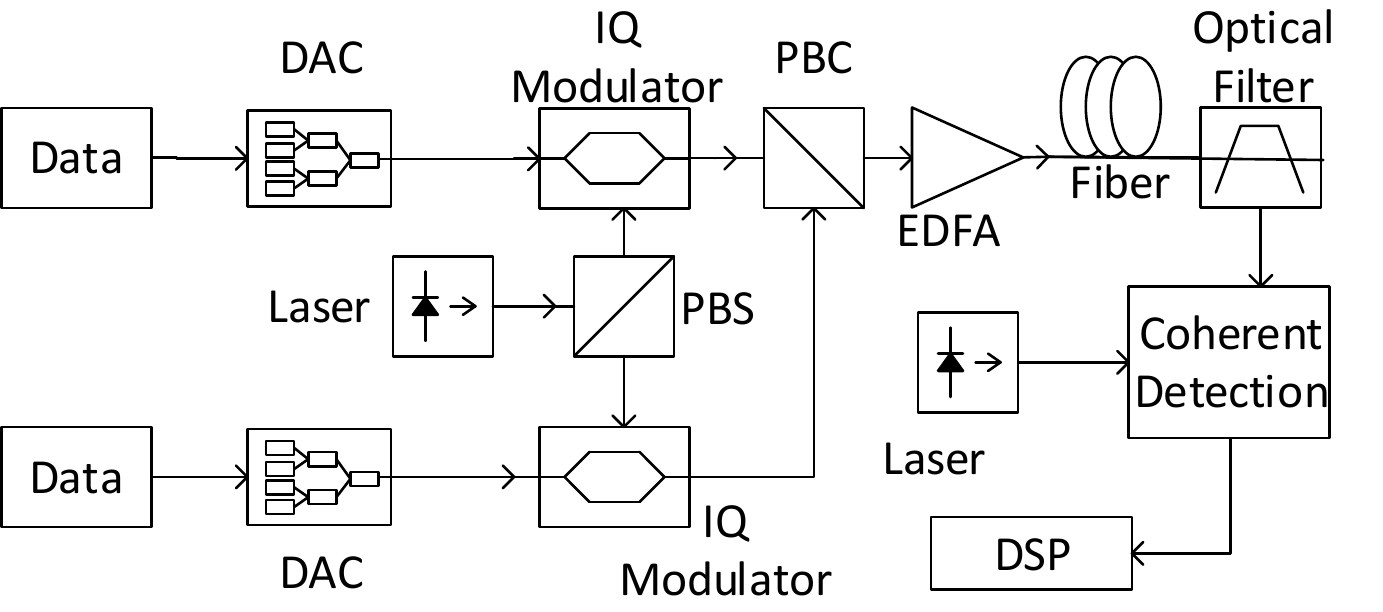}\caption{Simulation system setup. DAC: digital-to-analog converter; PBS: polarization beam splitter; PBC: polarization beam combiner; DSP: digital signal processing.}
\setlength{\belowcaptionskip}{-2.cm}
\end{figure}

\setlength\floatsep{-0.015cm}

Another important factor considered in MC is the number of samples needed to achieve a good performance\cite{bo2016modulation}. Fig. 7 shows the successful classification rate with different number of samples for 4-QAM, 8-QAM, 16-QAM, 32-QAM and 64-QAM when the OSNR equals to 12 dB, 15 dB, 19 dB, 22 dB and 24 dB, respectively. Note that we utilize the same OSNR values as in \cite{mai2017stokes}. To achieve 100\% successful classification rate for all of these five modulation formats, the proposed algorithm requires $K$ = 5500  samples, while $K$ = 8000 is needed for the algorithm in \cite{mai2017stokes}. More specifically,  the proposed algorithm needs 500, 3500, 5500, 5500 and 1000 samples to successfully classify 4-QAM, 8-QAM, 16-QAM, 32-QAM and 64-QAM, respectively. On the other hand, the algorithm in \cite{mai2017stokes} requires about 3000, 3000, 4000, 1000 and 8000 samples to successfully identify 4-QAM, 8-QAM, 16-QAM, 32-QAM and 64-QAM, respectively. The proposed algorithm  utilizes less samples to attain the same performance, which is a plus. Furthermore, the algorithm in \cite{mai2017stokes} employs a multi-dimensional tree to search among the neighbour points required for the clustering process, which has a complexity of $O(K\log{}K)$, while the proposed algorithm has a complexity of the order $O(K)$.
\begin{figure}[t]
\centering{}\includegraphics[width=0.8\columnwidth]{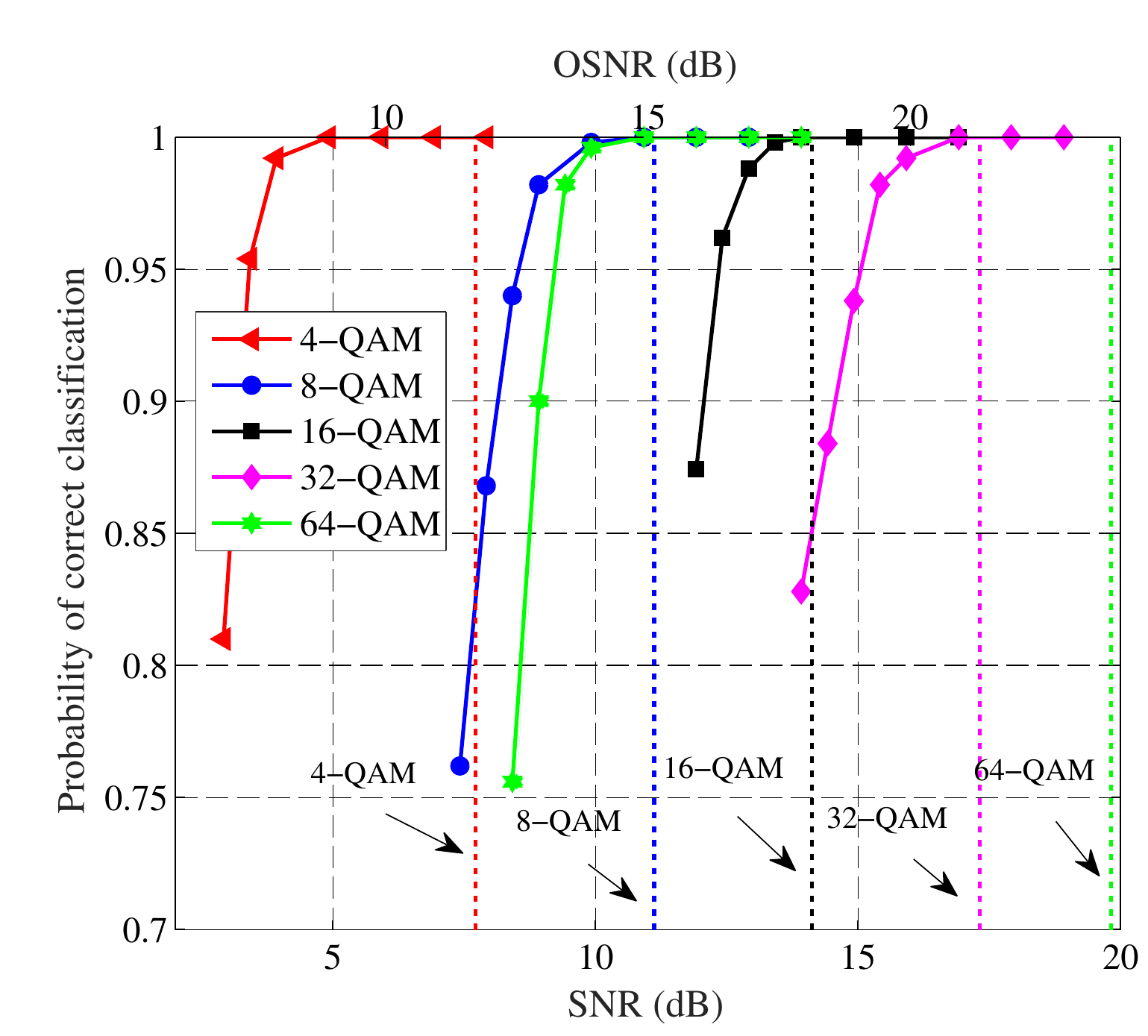}\caption{Probability of correct classification vs. SNR (bottom) and OSNR (top).}
\setlength{\belowcaptionskip}{-2.5cm}

\end{figure}
\setlength\textfloatsep{-0.001cm}

\begin{figure}
\begin{centering}
\includegraphics[width=0.8\columnwidth]{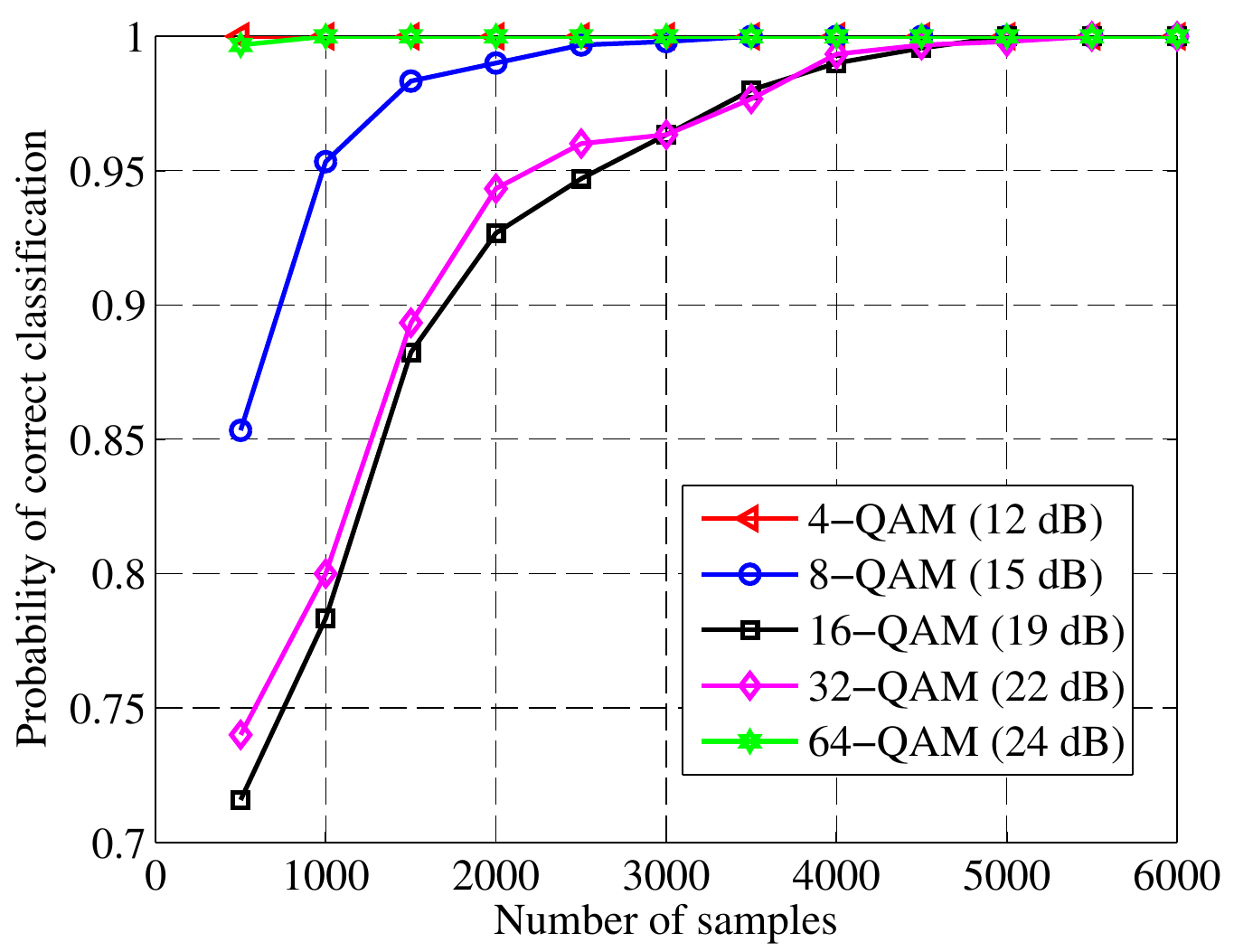}
\par\end{centering}
\centering{}\caption{Probability of correct classification vs. number of samples.}
\setlength{\belowcaptionskip}{-2.cm}
\end{figure}

\begin{figure}[t]   
\begin{centering}
\includegraphics[width=0.78\columnwidth]{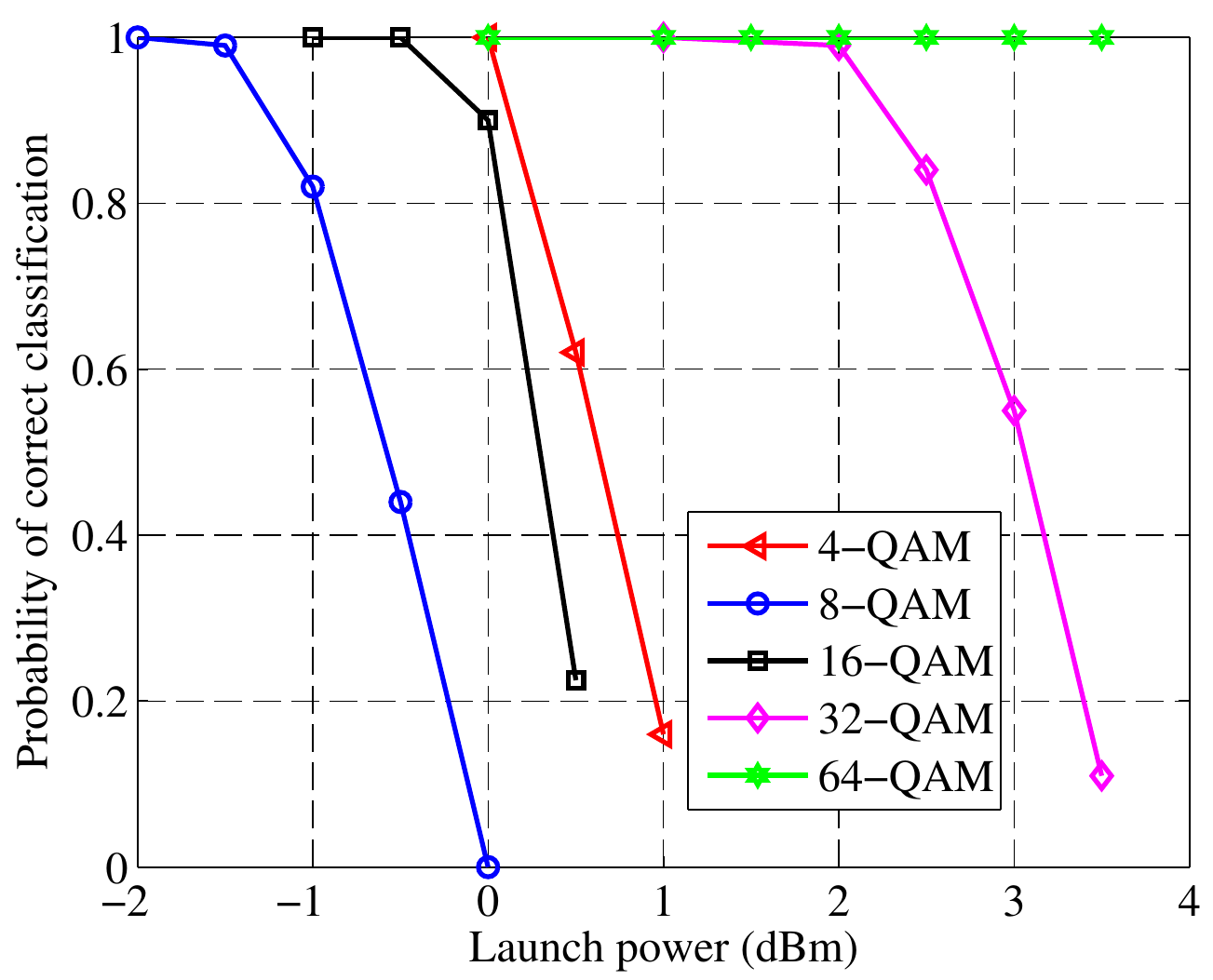}
\par\end{centering}
\caption{Probability of correct classification vs. launch power.}
\setlength{\belowcaptionskip}{-2.5cm}
\end{figure}

\vspace{-0.08cm}
The effect of fiber nonlinearity on MC is also investigated. The fiber nonlinearity is related to the launch power and fiber length, and it results in additional distortions that can become indistinguishable from the amplified spontaneous emission noise \cite{dong2012osnr}. We consider that the fiber lengths are 11200 km, 7600 km, 2800 km, 1040 km and 560 km for 4-QAM, 8-QAM, 16-QAM, 32-QAM and 64-QAM, respectively.  In this case, the BER for each modulation format is at the SD-FEC threshold when the launch power is 0 dBm. Then, the launch power is varied from -2 dBm to 3.5 dBm to see the impact of nonlinearity. According to Fig. 8, 64-QAM has 100\% successful classification due to the relatively short link length, while 8-QAM suffers mostly from the fiber nonlinearity. In general, results show that the modulation formats can be classified 100\% successfully when the launch power is below -2 dBm.

Additionally, the performance comparison is provided versus the MC scheme in \cite{liu2014modulation}, since this also utilizes the received signal's amplitude. When the number of samples is 10000, the required OSNRs for PDM 4-QAM, 16-QAM, 32-QAM and 64-QAM are listed in Table I. It can be seen that the proposed algorithm requires higher OSNR for 16-QAM only, while a lower OSNR is needed for 4-QAM, 32-QAM and especially 64-QAM when compared with the algorithm in \cite{liu2014modulation}. The complexity of the two algorithms is compared in Table II. It is assumed that $K$ samples are observed, and the number of bins for CDF plotting is 1000. It can be seen that both algorithms need only few real multiplications, while the proposed algorithm requires more real additions. However, it is worth noting that the extra processing time caused by the additions is not significant with high speed micro-processors. Note that the complexity for amplitude extraction is not considered here because both algorithms use features based on the signal's amplitude.

\begin{table}
\caption{Required OSNR (\textup{dB}) for 100\% successful classification.\label{tab:performance}}
\begin{centering}
\scalebox{0.8}{
\begin{tabular}{|c|c|c|c|c|}
\hline
Classification  algorithm  & 4-QAM & 16-QAM & 32-QAM & 64-QAM \tabularnewline 
\hline
\hline
Algorithm in [10] & $11.2$ & $16.5$ & $22$ & $24$ \tabularnewline
\hline
Proposed algorithm & $9$ & $18$ & $21$ & $15$ \tabularnewline
\hline
\end{tabular}}
\par\end{centering}
\end{table}

\begin{table}
\caption{Computational cost. \label{tab:Complexity2}}
\begin{centering}
\scalebox{0.8}{
\begin{tabular}{|c|c|c|}
\hline
Classification  algorithm  & Real multiplications & Real additions \tabularnewline 
\hline
\hline
Algorithm in [10] & $3$ & $4K$\tabularnewline
\hline
Proposed algorithm & 5 & $44K+4000$ \tabularnewline
\hline

\end{tabular}}
\par\end{centering}

\end{table}

\section{Conclusion}

We propose an MC algorithm based on the CDF of the received signal's normalized amplitude for coherent optical receivers. The proposed algorithm requires a fairly small number of samples and can be performed in the presence of the frequency offset and phase noise. Optical back-to-back experiments and extensive numerical simulations show that 100\% successful MC is achieved among PDM 4-QAM, 8-QAM, 16-QAM, 32-QAM and 64-QAM signals within the OSNR range of practical interests. Furthermore, the MC algorithm works well for launch powers
below -2 dBm. Such results are achieved with modest computational cost.


\vspace{-0.35cm}
\bibliographystyle{IEEEtran}
\bibliography{IEEEabrv,My_Ref_bib}

\begin{thebibliography}{10}
\providecommand{\url}[1]{#1}
\csname url@samestyle\endcsname
\providecommand{\newblock}{\relax}
\providecommand{\bibinfo}[2]{#2}
\providecommand{\BIBentrySTDinterwordspacing}{\spaceskip=0pt\relax}
\providecommand{\BIBentryALTinterwordstretchfactor}{4}
\providecommand{\BIBentryALTinterwordspacing}{\spaceskip=\fontdimen2\font plus
\BIBentryALTinterwordstretchfactor\fontdimen3\font minus
  \fontdimen4\font\relax}
\providecommand{\BIBforeignlanguage}[2]{{%
\expandafter\ifx\csname l@#1\endcsname\relax
\typeout{** WARNING: IEEEtran.bst: No hyphenation pattern has been}%
\typeout{** loaded for the language `#1'. Using the pattern for}%
\typeout{** the default language instead.}%
\else
\language=\csname l@#1\endcsname
\fi
#2}}
\providecommand{\BIBdecl}{\relax}
\BIBdecl

\bibitem{gerstel2012elastic}
O.~Gerstel, M.~Jinno, A.~Lord, and S.~B. Yoo, ``Elastic optical networking: A
  new dawn for the optical layer?'' \emph{IEEE Commun. Mag}, vol.~50, no.~2,
  pp. s12--s20, Feb. 2012.

\bibitem{Dobre2015signal}
O.~A. Dobre, ``Signal identification for emerging intelligent radios: Classical
  problems and new challenges,'' \emph{{IEEE} Instrum. Meas. Mag.}, vol.~18,
  no.~2, pp. 11--18, Apr. 2015.

\bibitem{dobre2007survey}
O.~A. Dobre, A.~Abdi, Y.~Bar-Ness, and W.~Su, ``Survey of automatic modulation
  classification techniques: classical approaches and new trends,'' \emph{IET
  Commun.}, vol.~1, no.~2, pp. 137--156, Apr. 2007.

\bibitem{dobre2005blind}
------, ``Blind modulation classification: a concept whose time has come,'' in
  \emph{Proc. IEEE SARNOFF Sympos.}, Apr. 2005, pp. 223--228.

\bibitem{khan2016modulation}
{F. N. Khan and et al.}, ``Modulation format identification in coherent
  receivers using deep machine learning,'' \emph{IEEE Photon. Technol. Lett.},
  vol.~28, no.~17, pp. 1886--1889, Sep. 2016.

\bibitem{borkowski2013stokes}
R.~Borkowski, D.~Zibar, A.~Caballero, V.~Arlunno, and I.~T. Monroy, ``Stokes
  space-based optical modulation format recognition for digital coherent
  receivers,'' \emph{IEEE Photon. Technol. Lett.}, vol.~25, no.~21, pp.
  2129--2132, Nov. 2013.

\bibitem{bo2016modulation}
T.~Bo, J.~Tang, and C.-K. Chan, ``Modulation format recognition for optical
  signals using connected component analysis,'' \emph{IEEE Photon. Technol.
  Lett.}, vol.~29, no.~1, pp. 11--14, Jan. 2017.

\bibitem{mai2017stokes}
X.~Mai, J.~Liu, X.~Wu, Q.~Zhang, C.~Guo, Y.~Yang, and Z.~Li, ``Stokes space
  modulation format classification based on non-iterative clustering algorithm
  for coherent optical receivers,'' \emph{Opt. Express}, vol.~25, no.~3, pp.
  2038--2050, Feb. 2017.

\bibitem{bilal2015blind}
S.~M. Bilal, G.~Bosco, Z.~Dong, A.~P.~T. Lau, and C.~Lu, ``Blind modulation
  format identification for digital coherent receivers,'' \emph{Opt. Express},
  vol.~23, no.~20, pp. 26\,769--26\,778, Oct. 2015.

\bibitem{liu2014modulation}
J.~Liu, Z.~Dong, K.~P. Zhong, A.~P.~T. Lau, C.~Lu, and Y.~Lu, ``Modulation
  format identification based on received signal power distributions for
  digital coherent receivers,'' in \emph{Proc. Opt. Fiber Commun. Conf.}, Mar.
  2014, pp. 1--3.

\bibitem{wang2016fold}
F.~Wang, O.~A. Dobre, C.~Chan, and J.~Zhang, ``Fold-based kolmogorov--smirnov
  modulation classifier,'' \emph{IEEE Signal Process. Lett.}, vol.~23, no.~7,
  pp. 1003--1007, Jul. 2016.

\bibitem{winzer2012high}
P.~J. Winzer, ``High-spectral-efficiency optical modulation formats,'' \emph{J.
  Lightw. Technol.}, vol.~30, no.~24, pp. 3824--3835, Dec. 2012.

\bibitem{ives2011estimating}
D.~J. Ives, B.~C. Thomsen, R.~Maher, and S.~J. Savory, ``Estimating osnr of
  equalised qpsk signals,'' \emph{Opt. Express}, vol.~19, no.~26, pp.
  B661--B666, Dec. 2011.

\bibitem{dong2012osnr}
Z.~Dong, A.~P.~T. Lau, and C.~Lu, ``Osnr monitoring for qpsk and 16-qam systems
  in presence of fiber nonlinearities for digital coherent receivers,''
  \emph{Opt. Express}, vol.~20, no.~17, pp. 19\,520--19\,534, Aug. 2012.

\end{thebibliography}


\end{document}